\documentclass{appolb}
\usepackage[dvips]{color}
\usepackage{ulem}
\usepackage{epsfig,epsf,subfig,graphicx,amsfonts,amssymb,color, subfig, color}
\usepackage{hyperref}

\begin{document}

\title{Multifractal Flexibly Detrended Fluctuation Analysis}
\author{R.~Rak$^{1}$, Pawe\l{}~Zi\c eba$^{1}$
\address{$^1$ Faculty of Mathematics and Natural Sciences, University of Rzesz\'ow, Pigonia 1, 35-310 Rzesz\'ow, Poland}}

\maketitle
\begin{abstract}

Multifractal time series analysis is a approach that shows the possible complexity of the system.
Nowadays, one of the most popular and the best methods for determining multifractal
characteristics is  Multifractal Detrended Fluctuation Analysis (MFDFA).
However, it has some drawback. One of its core elements is detrending of the series.
In the classical MFDFA a trend is estimated by fitting a polynomial of degree $m$ where $m=const$.
We propose that the degree $m$ of a polynomial was not constant ($m\neq const$)
and its selection was ruled by an established criterion.
Taking into account the above amendment, we examine the multifractal spectra
both for artificial and real-world mono- and the multifractal time series.
Unlike classical MFDFA method, obtained singularity spectra almost perfectly reflects the theoretical results
and for real time series we observe a significant right side shift of the spectrum.
\end{abstract}

\vskip0.5cm
{\it corresponding author; e-mail: rafalrak@ur.edu.pl}

\section{Introduction and motivation}
Since people attempted to understand the surrounding reality a lot of laws and methods have been found trying to describe this reality in a quantitative form. The major challenge for the current methods is to comprehend the behaviors and then attempt to model future states of time series, because people just have to deal with them in everyday life.
The most usual records of observable quantities in nature are in the form of time
series and their fractal and multifractals (nontrivial convolution of many fractals) properties have
been intimately investigated ~\cite{mandel1, kwapien2012}. There is a lot of evidence that this characteristic of empirical data coming from such diverse fields as physics of turbulence flows~\cite{muzy91}, geophysics~\cite {dimitriu00}, astrophysics~\cite{abramenko05}, physics of plasma~\cite {burlaga92}, physiology~\cite{ivanov99}, complex networks research~\cite{bianconi01} and econophysics~\cite{fisher97} is a very important feature of so-called complex systems. It seems to be that mono-(multi-) fractal effects (typical for complex systems) may come from the nonlinear
correlations as well as abundantly accompanying them the non-Gaussian heavy tails of fluctuations  or both equally~\cite{drozdz09}. There are many methods that can detect quantified the possible fractal nature of the data. For mono- and multi- fractal aspects, the most famous and recognized are: rescaled range (R/S) analysis~\cite{hurst1, mandel2, mandel3}, detrended fluctuation analysis (DFA)~\cite{peng1, hu1, grech1}, multi
fractal detrended fluctuation analysis (MFDFA) ~\cite{drozdz09, peng1, kantel1, ignaccolo1, oswiecimka06}, wavelet transform module maxima (WTMM) ~\cite{mol1, muzy1, muzy2, muzy4, struzik1, struzik2, oswiecimka06}, detrended moving average (DMA)~\cite{alessio1, carbone1, alvarez1, xu1} and  multifractal detrending moving average (MFDMA)~\cite{gu1}.
Nowadays there is also a lot of activity in the area of the so-called fractal cross-correlations: multifractal detrended cross-correlation analysis MFDXA~\cite{zhou1} and multifractal cross-correlation analysis (MFCCA)~\cite{oswiecimka14, horvatic11, podobnik11}.

\indent All the above mentioned methods of the type of $\backsim$DFA combine one expression, namely: \textit{detrending fluctuation} which for all types of $\backsim$DFA procedures can be briefly outlined as follows. Consider time series fluctuations ${x(i)}$ where ${i=1,...,N}$. First, for a given signal ${x(i)}$ the profile $X\left(j\right) =\sum_{i=1}^j[x_{i}-\langle x\rangle], {j=1,...,N}$ is calculated ($\langle ...\rangle$ denotes averaging over entire time series). Second, a signal profile is divided into $M_s=\lfloor N/s \rfloor $ disjoint segments $\nu$ of length
$s$. For each box $\nu$, the assumed trend is estimated by fitting a polynomial $P^{m}_{\nu}$ (where $m=const$ is an order of polynomial). Next, the trend is subtracted from the data. In general, we can use a variety of values of $m$ and the final result significantly depends on the value of $m$ ~\cite{oswiecimka13}. An example of detrending by different polynomials is shown in the Fig.1.
\begin{figure}[h!]
\begin{center}
\includegraphics[width=0.60\textwidth, height=0.41 \textwidth]{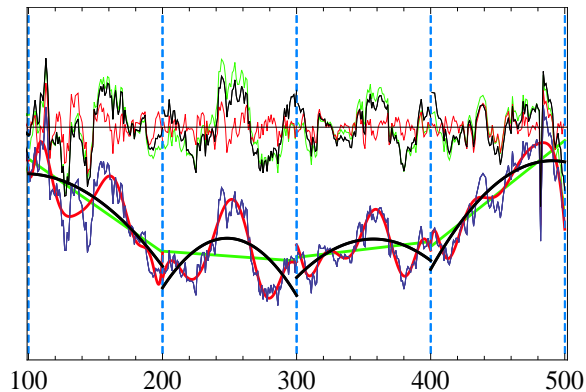}
\caption{An example of detrending data. A blue line denotes a time series before detrending. Green, black and red colours denote detrending polynomial $P^{m}_{\nu}$ (for $m=1$, $m=2$ and $m=10$) and detrended data respectively.}
\end{center}
\label{fig1}
\end{figure}

And here comes the question: what an order $m$ of polynomial to use? What polynomial best subtracts the trend? We cannot answer these questions unequivocally because we still do not know  what 'perfectly detrended data' means and what the detrending measure is.\\
\indent The MFFDFA method is tested on synthetic data (fractional Brownian motion and binomial multifractal cascade) and time series coming from real-world observables.

\section{MFFDFA algorithm}
As we mentioned above, Multifractal Flexibly Detrended Fluctuation Analysis (MFFDFA) has been developed on the basis the MFDFA algorithm ~\cite{kantel1}, wherefore a few steps are the analogous.

\indent \textit{Step 1}: Consider a signal ${x(i)}$ where ${i=1,...,N}$. For a given signal ${x(i)}$ the cumulative sum
\begin{equation}
Y\left(j\right) =\sum_{i=1}^j[x_{i}-\langle x\rangle],~~  {j=1,...,N}
\end{equation}
is calculated, where $\langle x \rangle$ denotes averaging over entire time series and $N$ is the length of time series.

\begin{figure}[h!]
\begin{center}
\includegraphics[width=0.65\textwidth, height=0.45 \textwidth]{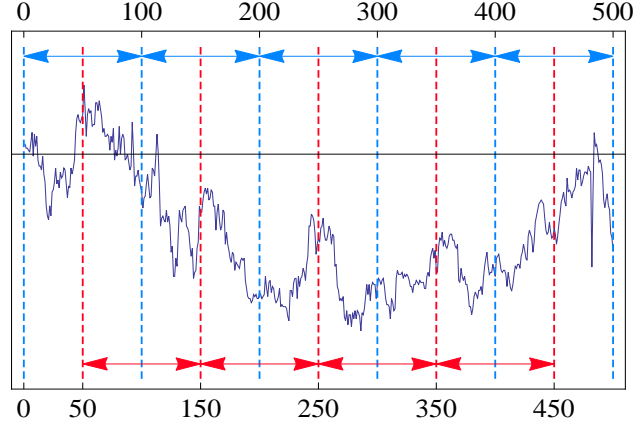}
\end{center}
\caption{An example of the division on segments $\nu$ having a length $s=100$. The situation is shown for $k=2$. The intervals overlap with a step $\lfloor s/k \rfloor=50$.}
\label{fig2}
\end{figure}

\indent \textit{Step 2}:
Then the profile $Y$ is divided into $M_{s_k}$ partly overlapping  segments $\nu$ of length $s$ with a step $ \lfloor s/k \rfloor$, where $k=1,2,3,...$. As a result of this modification we get approximately $k$ times more intervals $\nu$ of length $s$. Visualization of this idea is shown in the Fig.\ref{fig2}. It is visible, that for $k=1$ we obtain standard MFDFA method i.e. segments $\nu$'s non-overlapping. The minimum ($s_{min}$) and maximum scales ($s_{max}$) depend on the length $N$ of the time series under study. In practice, it is reasonable to take $s_{min}=30$ and $s_{max}=\lfloor N/10 \rfloor $.

\begin{figure}[h!]
\begin{center}
\includegraphics[width=0.65\textwidth, height=0.45 \textwidth]{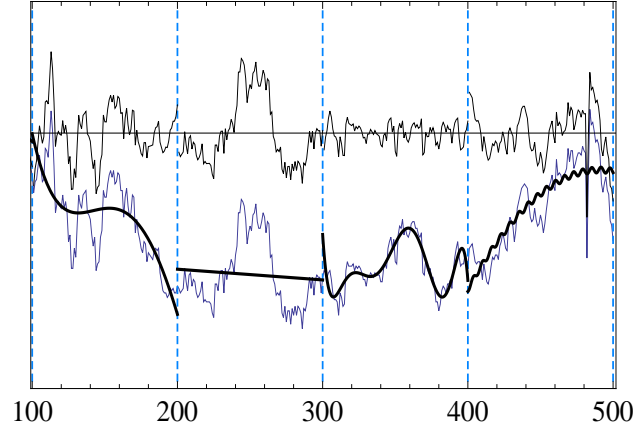}
\end{center}
\caption{An example of detrending data using different detrending functions.}
\label{fig3}
\end{figure}

\indent \textit{Step 3}:
For each box, the trend is estimated by fitting all functions of a set $Q=\{f_1,f_2,...,f_n\}$.
Next, for each segment only one detrended function $f_n$ is chosen of a set $Q$ in accordance with a prescribed criterion, which is then subtracted from the signal profile. An example of detrending data using various detrending functions is shown in the figure Fig.\ref{fig3}.

\indent \textit{Step 4}:

For the so-detrended signal a local variance $F^{2}(\nu,s)$ in each segment $\nu$ is calculated:
\begin{equation}
F^{2}(\nu,s)=\frac{1}{s}\Sigma_{k=1}^{s}\lbrace Y((\nu-1)s+k)-f^{(\nu)}_{n}(k)\rbrace^2 .
\end{equation}
Finally, $F^{2}(\nu,s)$ averaged over all $\nu$'s $q$th order fluctuation function is derived for all possible segment
lengths $s$:
\begin{equation}
F_q(s)=\lbrace\frac{1}{M_{s_k}}\Sigma_{\nu=1}^{M_{s_k}}[F^{2}(\nu,s)]^{q/2}\rbrace^{1/q},~~  q\in \mathbf{R}\setminus \{0\}
\label{Fq}
\end{equation}

In the case when $q=0$, the logarithmic version of Eq.~(\ref{Fq}) can be used~\cite{kantel1}:
\begin{equation}
F_{q=0}(s)=\frac{1}{M_{s_k}}\Sigma_{\nu=1}^{M_{s_k}}\ln|F^{2}(\nu,s)| .
\end{equation}

\indent \textit{Step 5}:
If the analysed signal is fractal, then $F_q$ scales within some range of $s$ according to a power law:
\begin{equation}
F_q\sim s^{h(q)},
\label{eq2}
\end{equation}
where $h(q)$ denotes the generalized Hurst exponent. For a monofractal signal, $h(q)$ is independent of $q$ ($h(q)=const$) and equals the Hurst exponent $h(q)=H$. On the contrary, for a multifractal time series, $h(q)$ is a decreasing function of $q$ ($h(q)\neq const$) and the simple Hurst exponent is obtained for $q=2$. The singularity  spectrum is calculated by means of the following relation:
\begin{equation}
\alpha = h(q)+qh^{'}(q) ~~~~ \hbox{and} ~~~~ f(\alpha)=q[\alpha-h(q)]+1 ,
\end{equation}

where $\alpha$ denotes the strength of a singularity spectrum and $f(\alpha)$ is the fractal dimension of a points set with particular $\alpha$. Typically, for multifractal data, the shape of the singularity spectrum is similar to a wide inverted parabola. The left and right wing of the parabola refers to the positive and negative values of $q$, respectively. The maximum of the spectrum is located at $\alpha(q=0)$. For a monofractal signal, the set representing $f(\alpha)$ reduces to a single point. The wealth of multifractality is evaluated by the width of its spectrum:
\begin{equation}
\Delta \alpha= \alpha_{max} - \alpha_{min},
\end{equation}
where $\alpha _{min}$ and $\alpha _{max}$ stand for the extreme values of $\alpha$. The richer is dynamics, the larger is $\Delta \alpha$ and the more developed is the multifractal.\\

~~For all tests, in this contribution, we choose arbitrarily a set of three fitting functions $Q=\{f_1,f_2,f_3\}=\{a x^2 + b x + c,~~ a \sin(x^2) + bx + c, ~~a x^3 + bx + c\}$ where $\{a,b,c\}$ are constant. The criterion of function selection was based on the so-called coefficient of determination $R^2$. For each box $\nu$, the function of set $Q$ will be selected with the highest value of $R$.

\section{Numerical tests of MFFDFA}
In order to investigate of MFFDFA method we carry out tests both for synthetic and the real-world data.
\subsection{Ordinary and fractional Brownian motion}
There are many different methods to create fractal time series i.a.: based on Fourier transform
filtering ~\cite{fBm1}, circulant embedding of the covariance matrix ~\cite{fBm2, fBm3}, midpoint displacement ~\cite{fBm4, fBm5}. In this contribution, we use the $Mathematica~~9.0$ to generate fractional Brownian motion ($fBm$). The long-term correlations of this Gaussian processes are completely characterized by the Hurst exponent $H\in (0,1)$. If $H=0.5$, time series is linearly uncorrelated and is the simplest case of a monofractal time series represented by the ordinary Brownian motion. For $0.5<H<1$, the data is persistent (positively correlated), which means that the signal more likely to follow the trend. If $0<H<0.5$, $fBm$ is antipersistent (negatively correlated) and consequently the signal has a tendency to change the trend direction.
\begin{figure}[h!]
\begin{center}
\hspace*{-.1in}
\includegraphics[width=0.85\textwidth, height=0.65 \textwidth]{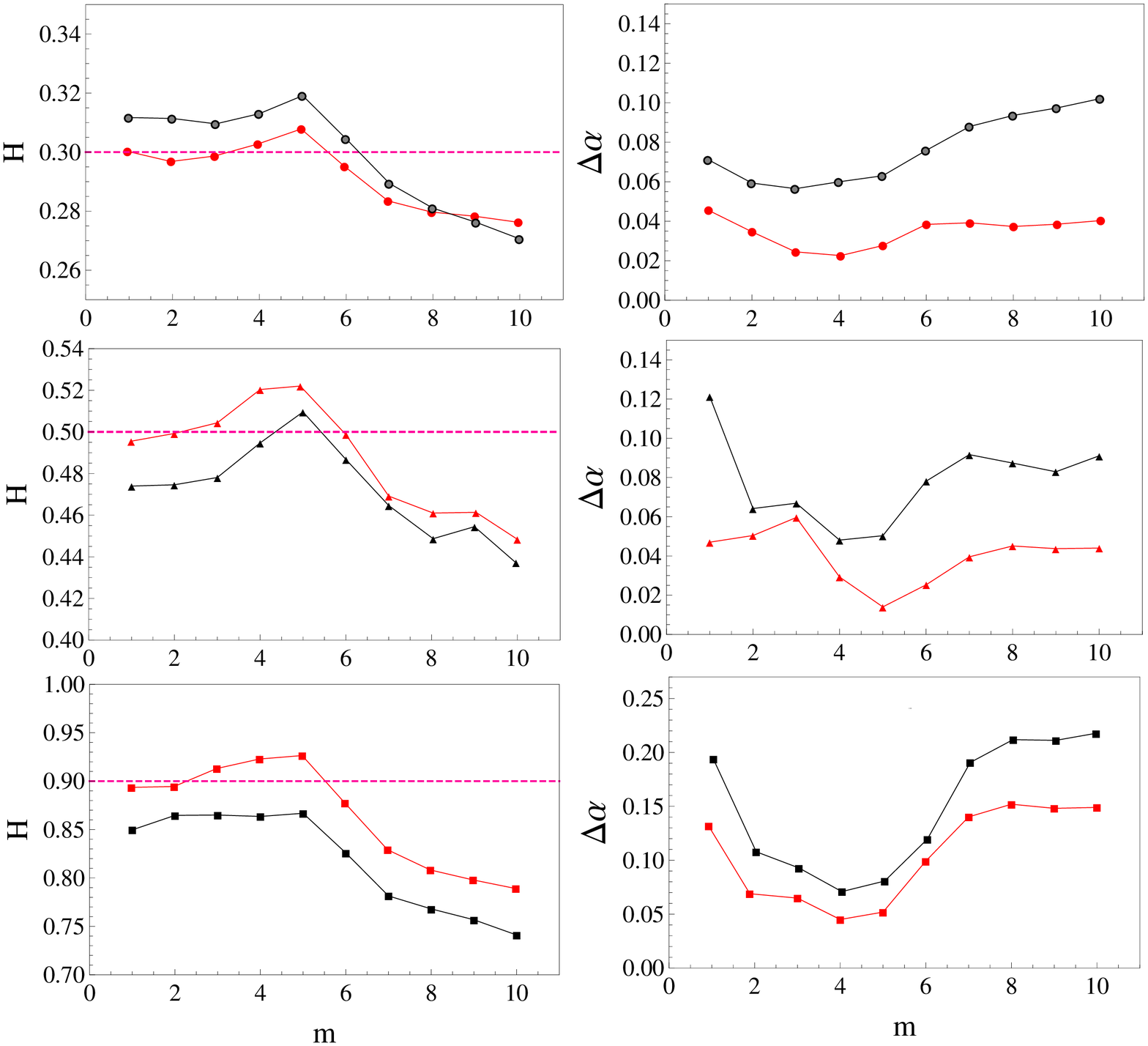}
\end{center}
\caption{Left panel: the average Hurst exponent $H$ as a function of detrending polynomial order $m$. Shows the comparative analysis of standard method MFDFA (black line) and modified MFDFA (red line) in accordance with the $Step~~2$ for three time series of $fBm$: H=\{0.3 (top), 0.5(middle), 0.9(bottom)\}. The right panel shows the average width of the singularity spectrum $\Delta \alpha$. The red dashed lines represent the theoretical values.}
\label{fig4}
\end{figure}
In our test, we have investigated $fBm$ with different Hurst indices $H=\{0.3, 0.5, 0.9\}$. In any case, we consider, relatively short, series of length of 10,000 points. The results for each process are averaged over its 10 independent realizations in order to be statistically significant. In addition, we restrict $q$ to $<-10,10>$ with a step 0.2 throughout this analysis.
In the figure \ref{fig4} was shown the comparative analysis of the standard method MFDFA and modified MFDFA in accordance with the $Step 2$. In both cases, the detrending polynomial of order $m$ in the range $<1,10>$ was used. It is clearly visible that the result strongly depends on the polynomial order $m$ - the higher the order of a polynomial, the results more deviates from the theoretical Hurst exponent $H$, both for standard as well as a modified version of the MFDFA method. However, a modified version of the MFDFA leads to a much closer result theory in particular for polynomials of order $m\leq5$. This is confirmed also by the analysis of the width of singularity spectrum $\Delta \alpha$ (right panel of figure \ref{fig4}), where the theoretical widths should be single points. In contrast to the standard method MFDFA the widths of spectrum are much closer to zero and results for $\Delta \alpha$ are, on average, about 70 percent better for modified MFDFA method.

\ \newline
Last test for monofractal time series we performed for the MFFDFA method. For the same time series as above i.e. for $H=\{0.3, 0.5, 0.9\}$ we received the following values $\{0.302, 0.497, 0.903\}$ respectively. What is more important, the analysis of the width of the singularity spectrum $\Delta \alpha$ indicates a much improved the accuracy of MFFDFA method - for successive values of $H$ we received the following values of $\Delta \alpha$: $\{0.01, 0.012, 0.015\}$. The results obtained are much closer to the theoretical results, i.e. are closer to zero and much better than the classic methods of MFDFA. This indicate that after the introduction of the amendments in step 2, 3 and 4 a effectiveness of the method significantly increased.

\subsection{Binomial Multifractal Cascade}
\begin{figure}[h!]
\begin{center}
\hspace*{-.1in}
\includegraphics[width=0.9\textwidth, height=0.45 \textwidth]{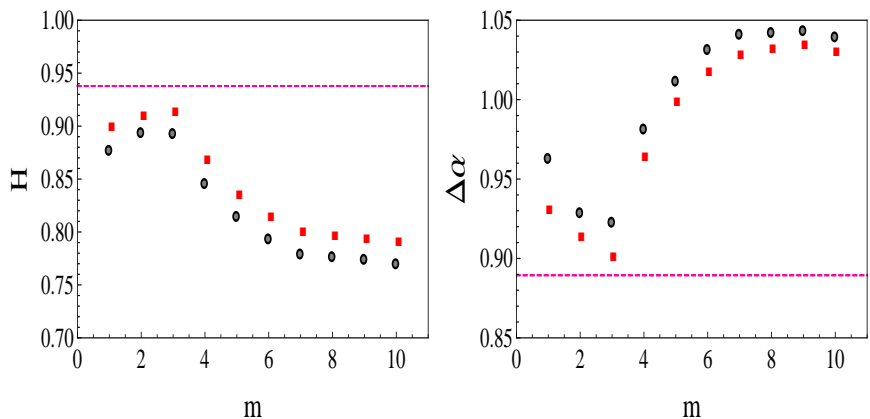}
\end{center}
\caption{Multifractal characteristics of the deterministic Binomial Cascade - the comparative analysis of standard method MFDFA (black points) and modified MFDFA (red points) in accordance with the $Step~~2$. Left panel: the Hurst exponent $H$ as a function of detrending polynomial order $m$. Right panel: the width of the singularity spectrum $\Delta \alpha$. The red dashed lines represent the theoretical values.}
\label{fig5}
\end{figure}

A well known example in the literature of the multifractal process is a binomial multiplicative cascade ~\cite{oswiecimka06}. This deterministic multifractal model can be defined by the following formula:
\begin{equation}
x_k=a^{n_{(k-1)}}(1-a)^{n_{max}-n_{(k-1)}},
\end{equation}
where $x_k$ is a time series of $2^{n_{max}}$ points $(k=1...2^{n_{max}})$, the parameter $a\in(0.5,1)$ is responsible for the fractal properties and $n_{(k)}$ denotes the number of 1's in the binary representation of the index $k$. The fractal properties of the model are quantified by the equations of the scaling exponent and the mutifractal spectrum:
\begin{equation}
\tau(q) = - {{- \ln [ a^q + (1-a)^q]} \over {\ln (2)}}
\label{binomialscaling}
\end{equation}

\begin{equation}
\alpha = - {1 \over \ln (2)} {{a^q\ln(a)+(1-a)^q\ln(1-a)} \over {a^q + (1-a)^q}}
\end{equation}

\begin{equation}
f(\alpha) = - {q \over \ln(2)} {{a^q\ln(a) + (1-a)^q\ln(1-a)} \over {a^q + (1-a)^q}}-{{-\ln[a^q+(1-a)^q]} \over {\ln(2)}}.
\end{equation}
To conduct the numerical analysis we create sets of time series of the same $n_{max}=17$. In this way, for a fixed value of $a$, we obtain a series 131072 points long. Unlike for the Brownian processes, due to the deterministic nature of the binomial cascades under study we create
only one time series. We restrict $q$ to $<-10,10>$ with a step 0.2 throughout below analysis.
In the figure  Fig.\ref{fig5} was shown the comparative analysis of standard method MFDFA (black points) and modified MFDFA in accordance with the $Step 2$ (red points). Here, assume that $a=0.65$.
For these tests we select a detrending polynomial of order $m$ in the range $<1,10>$. Relatively big values of $\Delta \alpha$ confirm that the analyzed time series has a multifractal nature. For all values of $m$, both for MFDFA and modified MFDFA, the estimated Hurst exponents are smaller than their theoretical counterparts. On the other hand, the $H$ index increases with $1<m\leq 3$, and for $4\leq m<10$, $H(m)$ is decreasing function of $m$. What is important, the most close results to theoretical one we observe for polynomials of order $m={1,2,3}$ and both for standard and modified methods MFDFA the polynomial of order 2 and 3 is the best approximation of the trend.

\begin{figure}[h!]
\begin{center}
\hspace*{-.1in}
\includegraphics[width=0.90\textwidth, height=0.35 \textwidth]{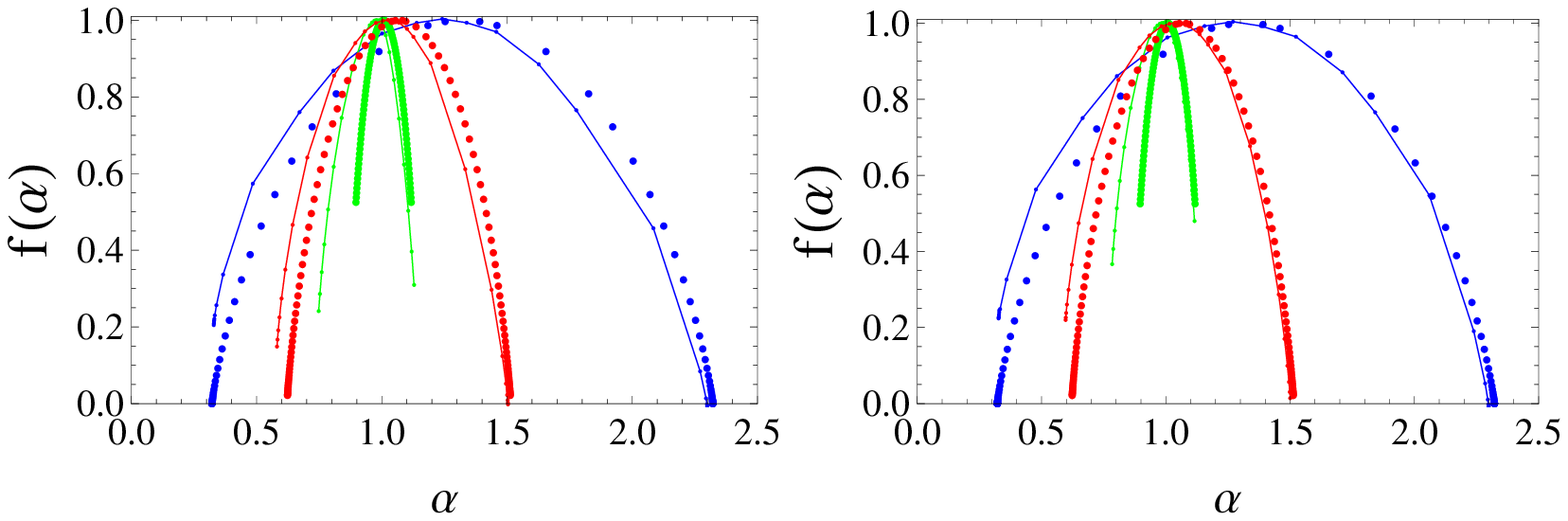}
\includegraphics[width=0.45\textwidth, height=0.35 \textwidth]{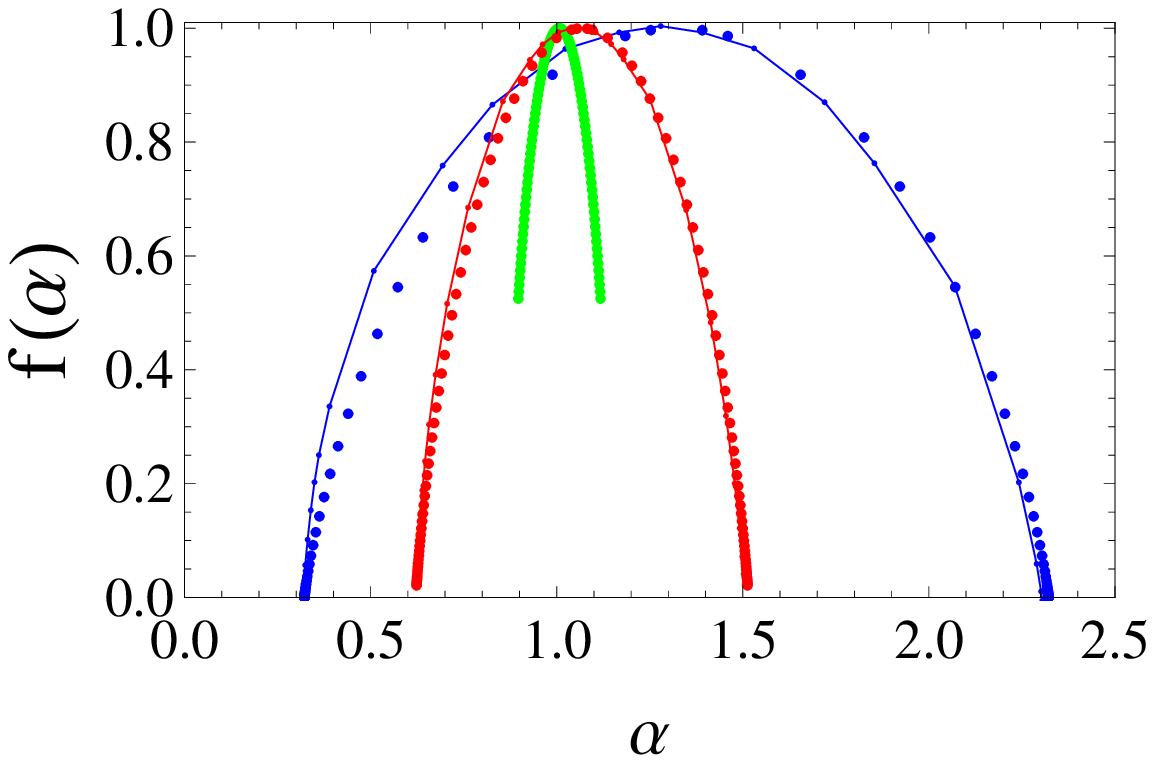}
\end{center}
\caption{Singularity spectra $f(\alpha)$ of binomial cascade for classical MFDFA and MFFDFA method. The dotted and the solid lines refer to the theoretical and the numerical results, respectively. Colors indicate results for various fractal parameters $a$ -- green, red and blue denotes $a=\{0.55, 0.65, 0.8\}$, respectively. Top left panel: analysis for standard MFDFA for detrending polynomial of order 3. Top right panel: analysis for modified MFDFA in accordance with the $Step~~2$. Bottom: analysis for MFFDFA method.}
\label{fig6}
\end{figure}

Another test what we do for synthetic multifractal time series is a comparative analysis of the standard MFDFA and MFFDFA method.
Here, we take 3 different values of $a=\{0.55, 0.65, 0.8\}$. Based on the above results, to the classical MFDFA method, we will use the detrending polynomial of order 3 (MFDFA3). In the case of the MFFDFA method we use set $Q$ of detrending functions and criterion of their selection described at the end of Section 2.
The results shown in Fig.\ref{fig6}. The most distant result from a theoretical we observe for the standard MFDFA method (top left panel). The better result was obtained for a modified version of this method (top right panel).
The result, which almost perfectly reflects the theory we have received in the case of the MFFDFA method (Fig.\ref{fig6}, bottom). In the present analysis, using the MFFDFA method, the functions from the set of $Q=\{a x^2 + b x + c,~~ a sin(x^2) + b x +c, ~~a x^3 + b x +c\}$ are selected on average $25\%$, $45\%$, $30\%$, respectively.
\begin{figure}[h!]
\begin{center}
\hspace*{-.1in}
\includegraphics[width=1\textwidth, height=0.42 \textwidth]{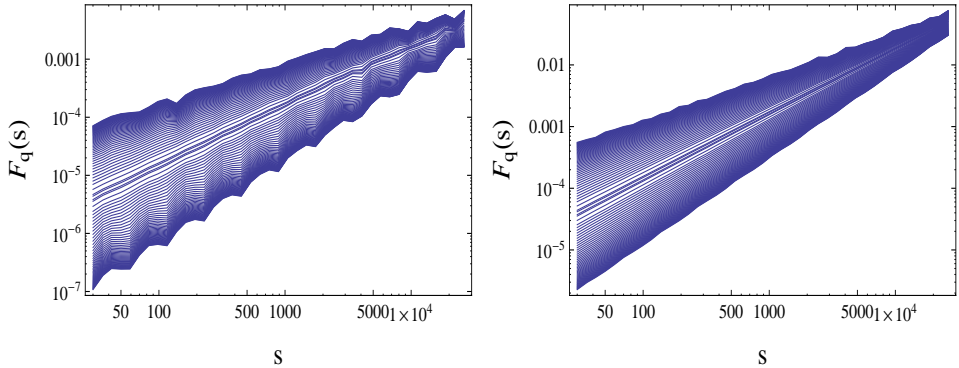}
\end{center}
\caption{Deterministic binomial cascade for $a=0.65$. Fluctuation function $F_q(s)$ calculated for standard MFDFA (left panel) and MFFDFA method (rigth panel). In both panels, the same parameters as in the top-left and bottom panel of Fig.\ref{fig6} were used.}
\label{fig7}
\end{figure}
What is important, the obtained result is a confirmation that the new MFFDFA method reflects very well not only monofractal but also multifractal nature of time series. Confirmation of this fact is the structure of the fluctuation function $F_q(s)$ shown in Fig.\ref{fig7}. It is clearly visible, that the fluctuation function of the new method MFFDFA is much more smooth and stable than a standard method MFDFA and this, in turn, leads to a result comparable with the theory (Fig.\ref{fig6}, bottom, the red graph).

Furthermore, there is the asymmetric nature of the singularity spectra (see Fig.\ref{fig6}). Discernible is primarily left-sided asymmetry. This kind of asymmetry results from distortions/depression of the large fluctuations. We think that this asymmetry is primarily due to imperfections detrending of time series. Obviously asymmetry is also caused by the statistical uncertainty of the MFDFA method. These types of effects are widely explained and modeled in ~\cite{drozdz2015}.
Similar calculations (as in Fig.\ref{fig6}) were also carried out using a polynomial of order 2. We found that the singularity spectra retained asymmetric nature and their width increased slightly (according to the results in Fig.\ref{fig5}).
\subsection{Real-world financial data}
Nowadays, it is believed, that the financial markets are one of the most complex systems all over the world. The huge number of individual transactions taken together, define very complex behaviour of the financial markets and lead to such characteristics of the financial data like multifractality, long memory, nonlinear correlations, the leverage effect, fat tails of financial data fluctuations, known together as the financial stylized facts~\cite{kwapien2012, arthur1999, ivanova1999, oswiecimka2005, oswiecimka2006, rak2005, plerou1999, drozdz2003, gabaix2003, rak2013}. These facts led us to choose this type of data in order to test the MFDFA and MFFDFA method.
We consider one-minute logarithmic price returns $r(i)=\ln(p(i+1))-\ln(p(i))$, representing dynamics of a sample US stocks -- Alcoa (AA), Walt Disney (DIS), Microsoft (MSFT) and Citigroup Inc. (C) being part of the Dow Jones Industrial index -- in the period 01-01-2008 -- 07-15-2011.
\begin{figure}[h!]
\begin{center}
\hspace*{-.1in}
\includegraphics[width=0.9\textwidth, height=0.45 \textwidth]{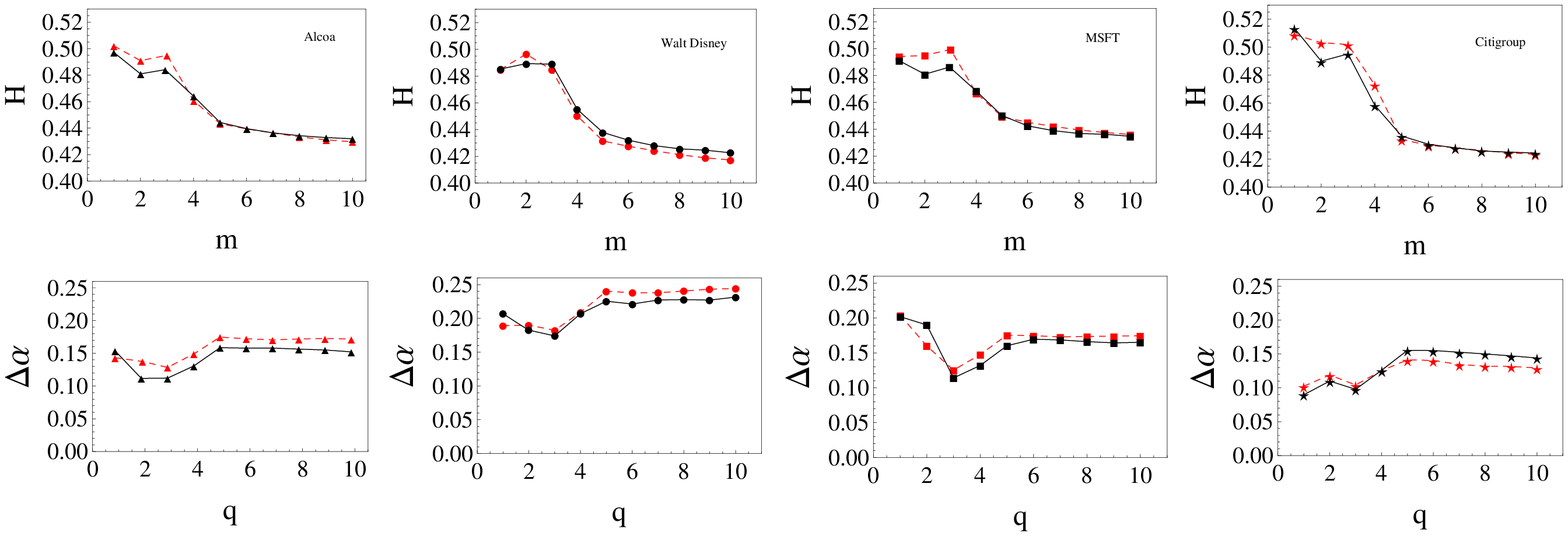}
\end{center}
\caption{Multifractal characteristics of the one-minute returns for stocks of Dow Jones Industrial index -- Alcoa, Walt Disney, Microsoft and Citigroup  (quoted over the period from 01-01-2008 to 07-15-2011) -- the comparative analysis of standard method MFDFA (black) and modified MFDFA (red) in accordance with the $Step~~2$. Top panels: the Hurst exponent $H$ as a function of detrending polynomial order $m$. Bottom panels: the width of the singularity spectrum $\Delta \alpha$ as a function $m$.}
\label{fig8}
\end{figure}
Moreover, these companies were selected from different industrial sectors. As above, we focus on the widths of the singularity spectra $\Delta \alpha$ and the Hurst exponent ($H(m)$) as a function of the number of the detrending polynomial of order $m$. Results for individual companies is shown in (Fig.\ref{fig8}). It is clearly evident that, as in the case of synthetic data, for small values ​​of $m$, the parameter $H$ is the highest and its value decreases with the increase of $m$.
This effect is seen both for the standard MFDFA (the black symbols on Fig.\ref{fig8}) and its modification (the red symbols). Moreover, for $m\geq4$, the value of $H$ is much less than $0.5$, which means that analyzed time series reveal high antipersitence. In this case, $H$ calculated for $m\leq3$ is approximately equal to 0.5, indicating uncorrelated data. In Figure \ref{fig8} (bottom panels), we present the estimated $\Delta \alpha$ as a function of $m$. Regardless of the industry, for all listed companies in both the width of the singularity spectrum $\Delta \alpha$ and the value of $H$ are similar in nature.

\begin{figure}[h!]
\begin{center}
\hspace*{-.1in}
\includegraphics[width=0.82\textwidth, height=0.45 \textwidth]{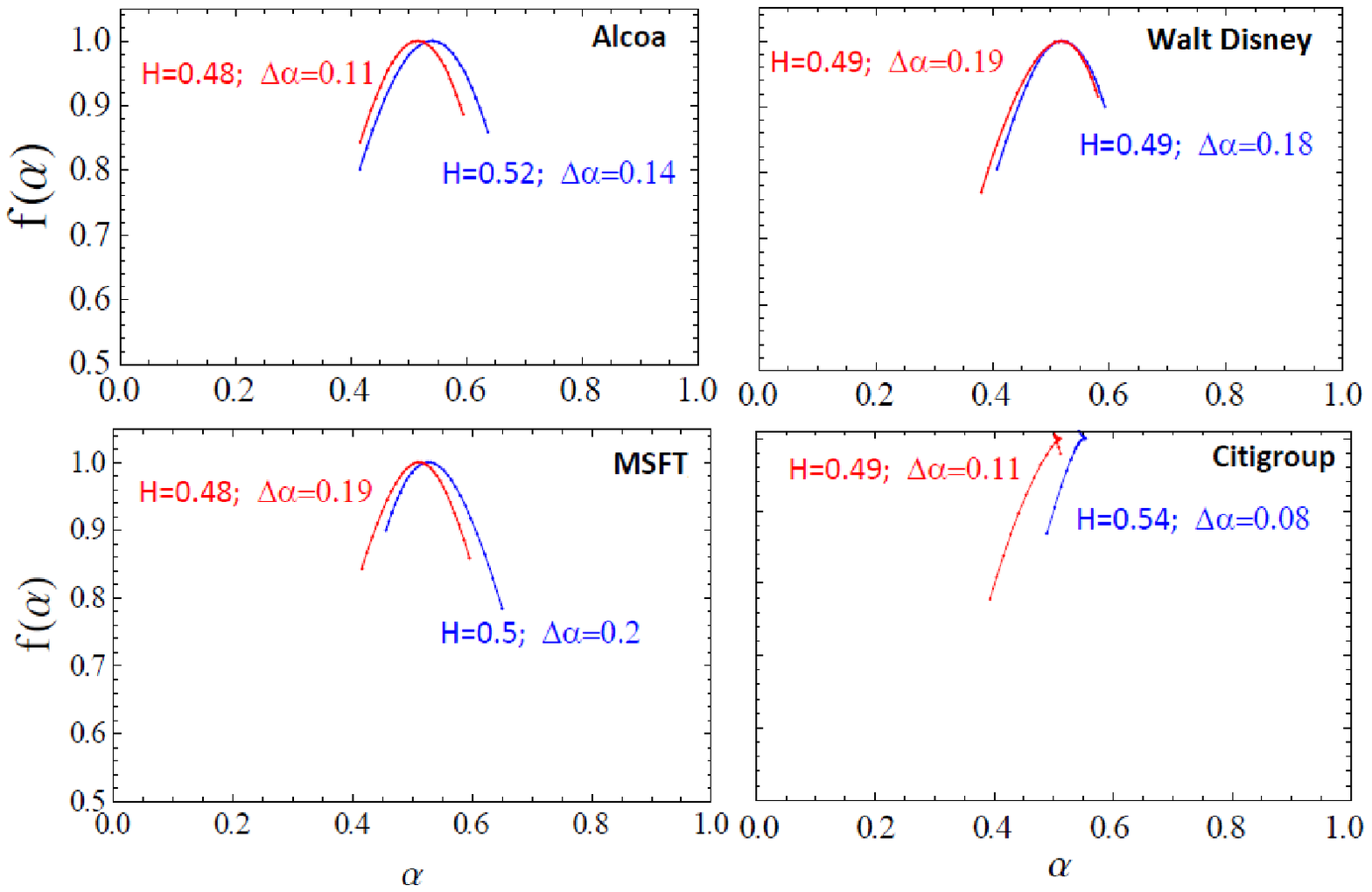}
\end{center}
\caption{Singularity spectra $f(\alpha)$ of the one-minute returns for 4 stocks of Dow Jones Industrial index quoted over the period from 01-01-2008 to 07-15-2011 -- the comparative analysis of MFDFA2 (red) and MFFDFA (blue) methods.}
\label{fig9}
\end{figure}

Complementing the above considerations is the comparative analysis of the two methods (MFDFA and MFFDFA). For the standard MFDFA method (in particular MFDFA2) we used here the polynomial of order 2 (the most common in the literature). Using the MFFDFA method, the detrending functions from the set of $Q$ are selected on average $21\%$, $40\%$, $39\%$, respectively. The results are shown in Fig.\ref{fig9}.
It is clear that, that the width of all spectra for both methods are similar. Interestingly, it is noted the differences between the values ​​of the Hurst exponent. Compared to the MFDFA method, for MFFDFA method, we observe a significant right side shift of the spectrum. This effect is most visible for Citigroup -- the difference between the values ​​of $H$ exceeds $10\%$. This in turn proves that the new MFFDFA method 'see the signal' as much more persistent.

\section{Summary}
In the present contribution we generalize the classical MFDFA method and we propose a novel theoretical algorithm - Multifractal Flexibly Detrended Fluctuation Analysis (MFFDFA) - that constitutes an extension of all methods of $\backsim$DFA type. We postulate that the degree $m$ of detrending polynomial was not constant in all segments $\nu$'s ($m\neq const$). In addition, the detrending process can be made using any function. Moreover, we choose the detrended function $f$ according to predetermined criteria. Obviously, the number and a type of functions $f$ of set $Q$ and the criterion for their selection is an open question.\\
In this paper we test this method for synthetic data and real-world data. It turns out that the MFFDFA method (for synthetic data) leads to significantly better results. For financial data we observe a shift of the multifractal spectra $f(\alpha)$ to the right which leads to the conclusion that the real-world data is actually more persistent than for the classical MFDFA method.

\section{Acknowledgements}
This work was partially supported by the Centre for Innovation and Transfer of Natural Sciences and Engineering Knowledge (University of Rzeszow).\\
The calculations were done at the Academic Computer Centre CYFRONET AGH,
Krak\'ow, Poland (Zeus Supercomputer) using Mathematica environment.

\end{document}